\documentclass[11pt]{article}
\usepackage{geometry}
\usepackage{amsmath}
\usepackage{amssymb,bm}
\usepackage{mathrsfs}
\usepackage{amsfonts}
\usepackage{bm}
\usepackage{graphicx}
\usepackage{endnotes}
\usepackage{array}
\usepackage{sectsty}
\usepackage{url}
\usepackage{mathpazo}
\usepackage{quoting}

\newcommand{\Sch}{Schr{\"o}dinger}

\begin{document}
\allsectionsfont{\normalsize}
\begin{center}
\large{\textbf{The Last Loophole in Bell’s Theorem?  \\
\smallskip
A \textit{Prima Facie} Problem With Superdeterminism\footnote{
Earlier versions of this paper were presented at meetings of the Western Canadian Philosophical Association for 2023, the Philosophy Colloquium at the University of Lethbridge in 2023, and the Canadian Philosophical Association for 2024.  Please direct comments or questions to \url{kent.peacock@uleth.ca}.  
}
}} 
\end{center}

\medskip
\begin{center}
\large{\textbf{Kent A. Peacock}} \\
\medskip
\textbf{Department of Philosophy, University of Lethbridge} \\
\end{center}

\bigskip\bigskip

%
\linespread{2.0}\selectfont\raggedright\parindent20pt


\begin{abstract}
 Hance and Hossenfelder recently claim that the extensive experimental confirmations of Bell's Theorem do not in fact demonstrate that nature is nonlocal, but merely that nature can be local only if the distant detector settings in a Bell-EPR setup are not statistically independent.  They suggest that such interdependence could take a local-deterministic form.  There is no question that in general quantum mechanics not only allows but demands that the distant detectors be entangled and thus correlated, even though it is possible to do experiments in which such correlations can be ignored.  The real question is whether entanglement itself could have a deterministic explanation.  We review why any such attempted deterministic underpinning for quantum statistics would run into immediate conflict with quantum indeterminacy and the noncommutativity upon which the latter is based.  Some cosmological implications are briefly explored.

\end{abstract}

\newpage

\section{Should the 2022 Nobel Prize be Rescinded?}
The 2022 Nobel Prize in physics was awarded to Alain Aspect, John F. Clauser, and Anton Zeilinger for their extensive studies of quantum mechanical entanglement and in particular their confirmation of Bell’s Theorem \cite{Bell64}, which has been called (by physicist H. P. Stapp) "the most profound result of modern science” \cite{Stapp75}.  J.S. Bell himself stated that according to his theorem, “maybe there must be something happening faster than light, although it pains me even to say that much” \cite{MC88}.  Whether an indication of faster than light effects should be an occasion for pain is an interesting question, but my main goal here is to respond to the recent claim by Hance and Hossenfelder that one important loophole remains in Bell’s momentous argument:
\begin{quote}
Contrary to what is often stated, these observations [for which Aspect et al. won their Nobel] do not demonstrate that “spooky action at a distance” is real and nature is therefore non-local.  Rather, the observations show that if nature is local, then statistical independence must be violated \cite{HH22}.
\end{quote}
By “statistical independence”, Hance and Hossenfelder mean the independence of detector settings from the same “hidden variables” in a Bell experiment that would determine the experimental outcomes.   Hidden variables are the presumptive inner machinery or coding built into the entangled particles at their common source which (Einstein had hoped) would be sufficient to explain the quantum correlations \cite{EPR35}.   Quantum correlations are typically functions of \emph{relative} detector parameters; for instance, in experiments involving correlated electron spins, the correlation coefficient is a sinusoidal function of the relative detector angle.  Bell assumed that the detector settings themselves are uncorrelated; in simple terms, this means that knowing the detector setting at one end of the apparatus tells us nothing about the angle chosen by the remote experimenter.  Bell himself was aware that the breakdown of this assumption could be a possible loophole in his argument \cite{Bell87,Wik01}.

As I will explain below in more detail, quantum mechanics certainly does allow for the possibility that the distant detector states (and thus their settings) could be entangled with each other and with the system being measured.  The usual assumption that they are not is merely a simplification in order that the entangled system of interest can be studied in isolation.  Thus, whether or not the distant settings can be correlated is not the real question, because in general quantum mechanics says they can---\emph{in a quantum mechanical way}.  The question is whether the statistical independence of the detectors can be violated in a \emph{deterministic} way.  Can all of the components of the system, including the detectors and the hypothetical hidden machinery internal to the entangled particles, be linked deterministically in just the right way to reproduce the predictions of quantum mechanics?  
Hance and Hossenfelder claim that this remains an open possibility.     

Suppose that everything in the universe is so thoroughly linked by causation that even though the experimenters \emph{feel as if} they can freely choose their detector settings, in fact all of the experimenters' choices, the detector settings, and the hidden variables within the particles themselves are thoroughly determined by some sort of common cause.  This apparent possibility is called \emph{superdeterminism} \cite{Hoss2020a}. 
One might have to postulate exotic sorts of classical causal links (such as wormholes or backwards-in-time influences), but some would feel that is an acceptable trade-off for the restoration of local determinism to physics \cite{Price96}.  

A loophole that gave pause to Bell himself is not lightly to be dismissed.  But it is not clear that the superdeterminists grasp how thorough-going a revision of modern physics would be needed in order to make their program work.  Bell’s Theorem is a special case of the so-called Bell-Kochen-Specker Theorem \cite{Mermin93,Bub97}, which states that in general there cannot be a Boolean (roughly speaking, set-theoretic) underpinning for quantum statistics.  The Bell-KS result itself is fundamentally a consequence of the fact that quantum observables come in noncommuting conjugate pairs.  As Jeffrey Bub  observes,
\begin{quote}
The really essential thing about a quantum world is the irreducible indeterminism associated with noncommutativity or non-Booleanity\dots \cite[p 240]{Bub97}.
\end{quote}
The deepest problem with superdeterminism, therefore,  is the \textit{prima facie} tension between determinism and noncommutativity.  
The superdeterminists must find a formulation of physics  that is entirely free of non-commutativity at a fundamental level,
 but which can also reproduce all of the well-verified predictions of quantum mechanics.  I will explain why this would be a very tall order indeed.  

\section{Entanglement}  
We must begin by saying something about what  entanglement is and where it comes from.  In quantum mechanics, if two or more systems have interacted dynamically (say they collided or decayed from a common source) then their combined system is described by a so-called tensor product of the states they would have if considered in isolation.  Such tensor product, or entangled, states have remarkable properties.  \Sch, who coined the term ``entanglement'', famously stated,
\begin{quote}
When two systems, of which we know the states by their respective representatives, enter into temporary physical interaction due to known forces between
them, and when after a time of mutual influence the systems separate again, then
they can no longer be described in the same way as before, viz. by endowing each
of them with a representative of its own. I would not call that \emph{one} but rather \emph{the}
characteristic trait of quantum mechanics, the one that enforces its entire
departure from classical lines of thought. \cite{Sch35a}
\end{quote}
All Bell's Theorem demonstrations involve correlations between measurements taken on particles belonging to entangled states.  

One can also speak of product states, systems which are presumed to act independently.  Product states can be correlated, but the correlations can be explained by common causes and will satisfy a Bell's Inequality (discused below).  The product state is a mathematical idealization.  Obviously, many systems can be treated as if they were product states 
but from the quantum point of view this simply means that their phase differences wash out in noise, thus yielding what we call classical mechanics. 
As Cohen-Tannoudji \textit{et al.} put it, 
\begin{quote}
 It can be shown\dots that an interaction between the two systems transforms an initial state which is a product into one which is no longer a product: any interaction between two systems therefore introduces, in general, correlations between them.  \dots  This question is very important since, in general, every physical system has interacted with others in the past \dots  \cite{CT77}
\end{quote}
If we take quantum mechanics seriously, and if we accept some version of the Big Bang cosmology in which all matter in the universe interacted at some point in the remote past, we must accept that \emph{all} physical systems in the universe are entangled.  It is entanglement all the way down.  This is a staggering thought that still has not been fully appreciated, although it has been obvious for quite some time.  

Why is there such a thing as entanglement?  Mathematically, it is a consequence of the superposition principle.  In quantum mechanics, physical systems are associated (in a mysterious way)  with linear objects called state vectors.  These behave very much like logarithms, in that they add up or superpose---any linear combination of allowed state vectors is an allowed state vector.  The tensor product state (here I cut a few mathematical corners) is simply the state of all possible linear combinations of the states of the subsystems.  

But why do we have to describe physical systems this way?  Whence linearity and the state vector?  As of this writing, no one knows.  The modern formalism of quantum mechanics was, in effect, reverse engineered from the phenomena, mostly in the period 1900--1927 \cite{Peacock2008}.  It was not the product of any philosophical agenda, but rather what nature forced on us---if you want good predictions, nature said, here is the mathematics you must use.  

So the distant detectors in a Bell experiment are entangled with the apparatus and with each other and so, of course, they are not statistically independent (though to a good approximation they can often be treated that way).  The real question is whether there could be a deterministic explanation for entanglement itself.

\section{For Whom the Modus Tollens}
Bell's Theorem (BT) is widely misunderstood, even by those who should be equipped to understand it.  One often hears something like the following response to BT:  ``Yes, distant particles in a nonfactorizable (entangled) state are correlated in ways that are difficult to explain classically, but don't worry, all of the information expressed by the correlations must have been built into the particles when they were emitted from their common source.'' The aim of Bell's argument was precisely to test this assertion, which he showed contradicts the predictions of quantum mechanics.  The violation of  locality is a corollary of this deeper result.  

The title of Bell's short but momentous paper of 1964 is, ``On the Einstein Podolsky Rosen Paradox.'' Bell was responding to a long debate launched by Einstein and co-authors in 1935 \cite{EPR35}.  EPR's argument is complex and subtle, but the gist of it can be expressed in simple terms.  EPR established a disjunction:  either the entangled particles have instructions built into them at the source that allows them to obey the statistical predictions of quantum mechanics, or a measurement choice made on one particle somehow influences the possible outcomes of measurements made on the other particle even after the particles are arbitrarily far apart in space or time.  Einstein thought it was beyond discussion that the latter suggestion was absurd; therefore, he insisted, the remote particle must have had a definite state before any measurements were made on its partner.  But since quantum mechanics (via the indeterminacy relations) apparently denies the possibility of a full specification of the state of the particles when they are emitted or at any other point in their histories, \emph{quantum mechanics must be an incomplete story about what goes on ``inside'' an elementary particle}.  


David Bohm \cite{Bohm51} advanced the discussion by describing a \emph{performable} version of the EPR thought experiment in terms of discrete spin measurements.  Bell realized that this opened the door to experimental tests of EPR's assumptions.  As Bell himself later described it \cite{Bern91}, Einstein had advocated what Bell called the \emph{genetic} interpretation of quantum correlations.  Siblings are more closely correlated in their heritable characteristics than people picked at random from the general population, but this demands no \textit{Spukhaftefernwirkungen} because  siblings share a common genetic heritage. The correlations between their physiological properties are built into their DNA.  Similarly, Einstein believed that the measurable properties of the distant particles in an EPR apparatus were somehow built into the particles at their common source.  They would be like the genetic properties shared by siblings.    What Bell showed in 1964 is that given the form of the typical correlations in entangled states, this is mathematically impossible: \emph{there is no quantum DNA}.  

There is an upper limit on how genetically correlated two or more persons can be, as seen in identical siblings.  Similarly, as Bell showed, there is an upper limit on how correlated two distant particles can be if their correlations depend entirely upon internal coding from their common source.  This limit is expressed by Bell's Inequality, which was shown by I.\ Pitowsky to be a special case of the logical ``conditions of possible [consistent] experience'' identified by George Boole in the 19th century \cite{Pit94}.   Bell-like inequalities express the assumption that the properties of the particles are Boolean, which (closely enough for our purposes here) means that the properties of the particles come in a package from their source and are not affected by the measurement process.  

Suppose Sue and Sonya are identical twins;  Sue is in Halifax and Sonya is in Vancouver.  They are having a phone conversation.  Even though they are as well correlated physiologically as two people can be, the informational content of their conversation (its mutual information)  could not have been entirely contained in the genetic information encoded in their DNA.  For instance, Sue might ask Sonya about the weather in Vancouver, and Sonya's reply (``It's raining'') depends partly on local conditions at the time of their conversation and could not have been encoded in the shared genetics they received from their parents at conception.  Are quantum correlations the same?  Are the distant particles, as it were, having a conversation about the remote detector settings (a conversation which, as Bell noted, would have to be faster than light)?  
As Bell himself noted (above), it is very difficult to interpret his result any other way.  However, most of the literature on this question 
consists of increasingly 
desperate attempts to do precisely this---to find any interpretation  of the Bell correlations whatsoever that would not involve some sort of nonlocal dynamics.

Bell's argument has the structure of a \textit{modus tollens}:  locality implies the Booleanity of particle properties; Booleanity implies the Bell Inequalities; both theory and observation show that the Bell Inequalities are violated; hence Booleanity is violated; hence locality is violated.  The superdeterminists don't deny that the Inequalities are violated, but they still hope to break the chain of consequences leading back to nonlocality.  

\section{Free Will is Not the Issue}
Whether human free will is consistent with physics is an important and interesting question, but it is not relevant here.  We can speak loosely and say that the measurement parameter choices can be made freely (or apparently so) by local observers 
at the remote measurement sites.  However, Bell experiments can be performed without any need for local parameter choices to be directly made by human experimenters.  For instance, one of the experiments for which Aspect won his Nobel was a delayed-choice experiment, in which local detector parameters are set automatically by a rapid randomizing device just before each particle enters its detector \cite{Aspect82b}.  No human intervention was required during the experiment.  What is crucial is that the choices are made by the randomizing switch \emph{after} the entangled particles are emitted from their sources and \emph{too late} for a light signal to travel from one detector to another.  So even though the correlation between the particles somehow ends up being a function of the relative angle between the detectors, information from one detector cannot have reached the other detector in time for it to have affected the measurement outcome---or can it?  \emph{That} is the conclusion that the superdeterminists wish above all else to avoid.

\section{Why Determinism Does Not Fare Well in a Quantum World}
Hossenfelder and Palmer define determinism as follows:
\begin{quote}
By deterministic we will mean that the dynamical law of the theory uniquely maps states at time $t$ to states at time $t^\prime$ for any $t$ and $t^\prime$.   \cite{Hoss2020a}
\end{quote}
The obvious problem with this notion is that the indeterminacy relations make it impossible to fully specify a physical state at a definite time.  If position could be known with full precision, then momentum would be completely uncertain, and vice versa.  Why can't we take it that the uncertainties in conjugate quantities such as position and momentum are merely indications of the practical difficulty of exact measurement, and that particles really do have definite values of these properties at all times even if it is not always possible to measure them with full precision?  For superdeterminism (or any sort of determinism) to work we would have to be able to do this.  

A bit of history will help.  Modern quantum mechanics was born in 1925 when Heisenberg, and shortly thereafter Dirac in a more general way \cite{Heis25,Dirac26}, grasped that the essential feature that distinguished classical physics from quantum physics was noncommutativity---the order in which measurement operations are performed makes a difference to the outcome. Heisenberg himself, as well as   \Sch\ and others, were quick to show that noncommutativity implies the Heisenberg Uncertainty Relations (more properly, Indeterminacy Relations) \cite{Wik02}.  

\Sch\ seems to have been the first to recognize that there is a compelling mathematical reason why the classical picture of measurement imprecision cannot explain the Indeterminacy Relations \cite{Sch35a}.  Here is an easy way of describing the problem \cite{Bub97}.  We can think of measurement as being like a process of interrogating the particles about their physical properties, and we can frame our questions so that they require yes/no answers  (``Is your spin-$z$ up?'').    For every possible property that the system could have, there will be a corresponding question.  Classically, the set of all possible such questions that could be posed to a given physical system would have a set of logically consistent answers (even if there were practical barriers to knowing some of those answers).  Classically, knowing the mass of a particle does not preclude knowing its velocity---one has simply gained more information about a given system.  

It does not work this comfortable way in quantum mechanics.  Instead, sharp answers to questions about (say) position will preclude sharp answers about momentum, in the sense that if we could imagine a set of all possible answers to all possible questions about the system, including questions about conjugate properties, that set would be contradictory.  As \Sch\ put it (somewhat elliptically) in 1935:
\begin{quote}
\dots if I wish to ascribe to the model at each moment a definite (merely not exactly known to me) state, or (which is the same) to \emph{all} determining parts definite (merely not exactly known to me) numerical values, then there is no supposition as to these numerical values \emph{to be imagined} that would not conflict with some portion of quantum mechanical assertions. \cite{Sch35b}
\end{quote}
The assumption that all possible quantum mechanical experimental questions have definite answers before we ask them is  \emph{logically contradictory} (or, in logician's terms, the set of all such propositions is not simultaneously satisfiable).   This result is the core of Bell's Theorem itself  \cite{Bub97}, and is built into the formalism of quantum mechanics.  In that language, asking for a particle to have simultaneous and exact values of position and momentum is like asking for a square circle.  Locality implies that the particles in an EPR-Bell scenario must have had a full set of sharp properties at their source.  Since some of those properties fail to commute, there cannot be a full set of answers to all possible experimental questions about those properties.  

Because the kind of classical determinism sought by Hossenfelder and others requires a way of completely and \emph{consistently} specifying the state of a system at a given time, it seems to be, \textit{prima facie} at least, ruled out of court by the mathematical structure of quantum mechanics---unless the superdeterminists can find a fully deterministic underpinning for noncommutativity itself.

\section{The Bigger Picture}  




This story has cosmological implications.  The superdeterminists 
think they can beat Bell by expanding the location of the Boolean set of local properties beyond the envelope of the entangled particles themselves in the Bell experiment.  Superdeterminism is, in effect, Einstein's genetic interpretation writ large, with the common cause becoming literally the whole universe.  However, if it is entanglement all the way down, it must be entanglement all the way up.  If quantum mechanics really is as fundamental as it seems to be, the universe is a quantum object and \emph{could have} no exact classical-local-deterministic description.  As W.\ Demopoulos argued \cite{Dem04}, any thinkable description of a quantum system is necessarily incompletable, and that must include the universe itself .  The best that the would-be classical determinist could hope for would be an indefinitely large set of convenient semi-classical approximations, such as the approximation discussed earlier in this paper in which we take the distant detectors in a Bell-EPR experiment to be statistically independent.   

Any deterministic or superdeterministic attempt to find a Boolean story underpinning quantum statistics is bound to fail, because what matters is the presumed Booleanity of the common causes, not their distribution in space and time, or how many exotic ways the hidden machinery might be connected.   If I am right, therefore, superdeterminism will be refuted by a suitable generalization of Bell's Theorem itself.  But that task remains to be done in a precise and convincing manner.  

Behind all of this lurks a deeper question:  whither the quantum of action that forces commutators to be non-zero?  Is it purely contingent---that is, did it simply happen to come out to the value that it has for essentially historical reasons, the way that some individual person's weight just happens to be (say) 71 kg?  Or is it the consequence of deep logical and/or mathematical constraints that we do not yet understand?  The universality of Planck's constant suggests the latter, and if this is correct then there can be \emph{no question} of a fully deterministic underpinning for quantum statistics, super- or otherwise.  But this piece of very important homework remains to be done.


%

\end{document}